\documentclass[epj]{svjour}
\usepackage{graphics}
\begin{document}
\title{Don't forget to measure $\Delta s$}
\author{Stephen F. Pate}
\institute{Physics Department, New Mexico State University, Las Cruces NM 88003}
\date{Received: date / Revised version: date}
%
\abstract{
This talk explores our lack of knowledge of the strange quark contribution to the nucleon
spin, $\Delta s$.
Data on $\Delta s$ from inclusive and semi-inclusive polarized deep-inelastic scattering 
will be reviewed, followed by a discussion of how the ongoing program of parity-violating 
elastic electron-nucleon scattering experiments, that seek out the strange electromagnetic 
form factors of the nucleon, need to have an estimate for the strange axial form factor 
to carry out that program, and how the value of $\Delta s$ extracted from the 
DIS experiments has filled that role.  It is shown that elastic $\nu p$, 
$\bar{\nu} p$, and parity-violating $\vec{e}p$ data can be combined to extract 
the strange electric, magnetic $and$ axial form factors simultaneously.
A proposed experiment that could address this important issue if briefly previewed.
} 
\maketitle

\section{Introduction}

In the current experimental program of nucleon structure studies, we find two broad areas
of experimentation.  On the one hand, elastic scattering of electrons from nucleons 
is used to measure the electromagnetic and axial form factors of the nucleon, over a 
range of momentum transfer of $0.1 < Q^2 < 10$ GeV$^2$.  These experiments have taken 
place at a variety of laboratories over the
years, with the current program focused at MIT-Bates, JLab, and
Mainz.  One of the highlights of the current program is the emphasis on nailing down the
strange quark contributions to the electromagnetic form factors, through the exploitation
of the interference between photon and $Z$-boson exchange processes.  On the other hand, 
deep-inelastic scattering of muons and electrons from nucleon and nuclear targets, 
historically responsible for the discovery of the partonic structure of matter, 
continues to play a role in the exploration of the distribution of quarks and gluons in
nucleons.  One of the highlights here is the focus, over the last 15 years, 
on the spin structure of the nucleon.  The deep-inelastic exploration of nucleon
spin takes place now at both leptonic and hadronic facilities, the spin program at 
RHIC being the most notable example of an hadronic facility taking on this physics topic.

QCD provides a simple framework in which these two experimental programs are joined
together.  The asymmetries observed in the polarized deep-inelastic scattering experiments
arise from the antisymmetric part of the virtual Compton amplitude, which contains at
its heart the nucleon axial current, $\bar{q}\gamma_\mu \gamma_5 q$.  
In the quark-parton model, inclusive scattering of leptons from nucleon targets measures 
the nucleon structure function $F_1$,
$$F_1(x)=\frac{1}{2}\sum_q e_q^2 q(x)$$
where $e_q$ and $q(x)$ are respectively the charge and parton distribution function for 
quarks of flavor $q$.
Inclusive scattering of {\bf polarized} leptons from {\bf polarized} nucleon targets 
measures the 
{\bf spin-dependent} nucleon structure function $g_1$,
$$g_1(x)=\frac{1}{2}\sum_q e_q^2 \Delta q(x)$$
where now $\Delta q(x)$ is a polarized parton distribution function.  In QCD, these
distribution functions take on a scale dependence: $\Delta q(x,Q^2)$.
At the same time,
the axial form factors $G_A^q(Q^2)$ measured in elastic scattering are themselves 
matrix elements of the axial current,
$$\raisebox{-.4ex}{${}_N$}\!\left<p'|\bar{q}\gamma_\mu \gamma_5 q|p\right>_N 
            = \bar{u}(p')\gamma_\mu \gamma_5 G_A^q(Q^2)u(p)$$
where the matrix element has been taken between two nucleon states of momenta $p$ and $p'$,
and $Q^2 = -(p'-p)^2$. The diagonal matrix elements of the axial current are called the
axial charges,
$$\raisebox{-.4ex}{${}_N$}\!\left<p|\bar{q}\gamma_\mu \gamma_5 q|p\right>_N = 2Ms_\mu \Delta q$$
where $M$ and $s_\mu$ are respectively the mass and spin vector of the nucleon.  The
quantities $\Delta q$ are called ``axial charges'' because they are the value of the 
axial form factors at $Q^2=0$;  that is to say, for example, $G_A^s(Q^2=0) = \Delta s$.
The connection between the two sets of observables lies in a well-known QCD sum rule for
the axial current, namely that the value of the axial form factor at $Q^2=0$ is equal to the
integral over the polarized parton distribution function measured at  $Q^2=\infty$.  
For example,
$$\Delta s = G_A^s(Q^2=0) = \int_0^1 \Delta s(x,Q^2=\infty) dx.$$
In this way, the axial charges $\Delta q$ provide the link between the low-energy 
elastic scattering measurements of axial form factors and the high-energy deep-inelastic
measurements of polarized parton distribution functions.

Of course, there are practical difficulties in the full exploration of this sum rule.
No scattering experiment can reach $Q^2=0$ or $Q^2=\infty$, and no deep-inelastic
experiment can ever reach $x=0$.  However, the consequences of these difficulties are more
severe in some cases than in others.  Our inability to reach $Q^2=\infty$ in the deep-inelastic
program means that QCD corrections enter into the sum rule written above, and there is
much theoretical experience in calculating these corrections.  
While the low-energy elastic experiments
cannot reach $Q^2=0$, there is not expected any divergent behaviour of the form factors near
$Q^2=0$ and so the idea of extrapolating to $Q^2=0$ from measurements at low, non-zero 
$Q^2$ is not met with alarm.  On the other hand, the limitations imposed by our inability
to reach $x=0$ in the deep-inelastic experiments are more problematic.  The unpolarized 
parton distribution
functions $q(x)$ are all known to increase rapidly as $x\rightarrow 0$ and there is no calculation
of the expected behaviour near $x=0$ to rely upon for an extrapolation from measurements
made at $x \neq 0$.  Similar comments apply to the polarized parton distributions $\Delta q(x)$.
Unpolarlized measurements of the parton distributions at HERA have reached very low values of
$x$, nearing $x=3\times10^{-5}$, but the corresponding measurements of the polarized distribution
functions, from data at SLAC, CERN, and DESY, only reach $x=3\times 10^{-3}$.
Therefore, measurements of the axial charges place important constraints on the behaviour of
the distributions $\Delta q(x)$ in the unmeasureable low-$x$ region.

\section{$\Delta s$ from Inclusive Leptonic Deep-inelastic Scattering}

As mentioned earlier, the double-spin asymmetries in polarized inclusive leptonic deep-inelastic scattering measure the
spin-dependent nucleon structure function $g_1$:
$$g_1(x)=\frac{1}{2}\sum_q e_q^2 \Delta q(x).$$
In leading order QCD, these functions take on a scale dependence:
$$g_1(x,Q^2)=\frac{1}{2}\sum_q e_q^2 \Delta q(x,Q^2).$$
In next-to-leading order (NLO) QCD, there are significant radiative corrections and the relation between
$g_1$ and the $\Delta q$ becomes more complex.  In the discussion here, we will limit our attention to the leading-order QCD
analysis because the NLO version of the analysis does not change the result (nor the uncertainty) for $\Delta s$ very much,
and the problems to be pointed out exist at all orders, because they are problems coming from the data itself.

We will use the analysis from the SMC Collaboration\cite{SMC1997} as a model. They measured $g_1(x,Q^2)$ over a wide
kinematic range, $0.003<x<0.70$ and $1.3<Q^2<58.0$.  This coverage is not a rectangle, {\em i.e.} there is a correlation between
$x$ and $Q^2$ in the accepetance of the experiment, and so for a reasonable analysis it is
necessary to use QCD to eveolve all the data to a single value of $Q^2$,
in this case $Q^2=10$~GeV$^2$.  In the process of performing this evolution, a fit function for $g_1$ is produced.  Then, to
integrate the distibution $g_1$ over $0<x<1$, it is necessary to extrapolate to $x=1$ and $x=0$.  The extrapolation to $x=1$
makes use of the fact that $g_1$, being a difference of two quark distributions, must go to 0 as $x\rightarrow 0$.  This 
requirement is satisfied in this analysis by assuming the measured experimental asymmetry to be constant for $x>0.7$.
The extrapolation to $x=0$, on the other hand, is not straightforward, as the expected 
behaviour of $g_1(x)$ for $x\rightarrow 0$ is unknown.  In this analysis, two methods were used. In one, the QCD evolution
fit was simply extrapolated to $x=0$. In another, called the ``Regge extrapolation,'' the value of $g_1$ was assumed to be
constant for $x<0.003$.  The two values of
the integral of $g_1$ from these two extrapolations are
\begin{eqnarray*}
\Gamma_1 = \int_0^1 g_1(x)dx &=& 0.142\pm0.017~~~\mbox{``Regge''}\\
                             &=& 0.130\pm0.017~~~\mbox{QCD fit}.
\end{eqnarray*}
This integral is related to the axial charges:
\begin{eqnarray*}
\Gamma_1 & = & \int_0^1 g_1(x)dx = \frac{1}{2}\sum_q e_q^2 \int_0^1 \Delta q(x)dx \\
         & = & \frac{1}{2}\left[\frac{4}{9}\Delta u + \frac{1}{9}\Delta d + \frac{1}{9}\Delta s \right].
\end{eqnarray*}
Now, assuming that SU(3)$_f$ is a valid symmetry of the baryon octet, and
using hyperon $\beta$ decay data, then two other relations between the three axial charges are determined:
$$\Delta u - \Delta d = g_A = F+D \mbox{~~~~and~~~~}\Delta u + \Delta d - 2\Delta s = 3F-D$$
where $g_A = 1.2601\pm0.0025$ and $F/D = 0.575\pm0.016$ (in 1997).  Now one may solve for the axial charges, and
the results are shown in Table~\ref{SMC_table}.
\begin{table}[h]
\caption{\label{SMC_table} Results for the axial charges from the SMC analysis\protect\cite{SMC1997} 
of their inclusive DIS data.}
\begin{center}
\begin{tabular}{c|c|c}
\hline\hline
      & ``Regge'' & QCD fit \\
\hline
$\Delta u$ & $~~0.84\pm 0.06$ & $~~0.80\pm 0.06$ \\ 
$\Delta d$ & $ -0.42\pm 0.06$ & $ -0.46\pm 0.06$ \\
$\Delta s$ & $ -0.08\pm 0.06$ & $ -0.12\pm 0.06$ \\
\hline\hline
\end{tabular}
\end{center}
\end{table}
It is well to note that, of course, the error bars quoted here do not include
any estimate of the theoretical uncertainty
underlying the assumption of SU(3)$_f$ symmetry.  They do include an estimate
of the uncertainty due to the extrapolations, but
of course that is only an estimate because the actual behaviour of $g_1$ is
unknown in the $x\rightarrow 0$ region.  The only
conclusion to be drawn for $\Delta s$ from this analysis is that it may be
negative, with a value anywhere in the 
range 0 to $-0.2$.

\section{$\Delta s(x)$ from Semi-inclusive Leptonic Deep-inelastic Scattering}

In semi-inclusive deep-inelastic scattering experiments, a leading hadron is observed in coincidence with the scattered lepton,
allowing a statistical identification of the struck quark, and thus a measurement of the $x$-dependence of the individual
$\Delta q(x)$ distributions.  (Inclusive scattering only measures the total structure function $g_1(x)$.)
The HERMES Experiment \cite{Ackerstaff:1998av} at DESY was especially designed to make this measurement.
HEMRES measured double-spin asymmetries in the production of charged hadrons in polarized deep-inelastic scattering of positrons
from polarized targets; specifically, the asymmetry in the production of charged pions on targets of hydrogen and deuterium, 
and of charged kaons in scattering from deuterium.  There is no assumption of SU(3)$_f$ symmetry in their analysis.  
They extract the following quark polarization distributions, over the range $0.023<x<0.30$\cite{Airapetian:2003ct}:
$$\frac{\Delta u}{u}(x)~~~\frac{\Delta d}{d}(x)~~~\frac{\Delta \bar{u}}{\bar{u}}(x)~~~\frac{\Delta \bar{d}}{\bar{d}}(x)~~~\frac{\Delta s}{s}(x)$$
where $\frac{\Delta s}{s}(x)$ is defined to be the sum of $\frac{\Delta s}{s}(x)$ and $\frac{\Delta \bar{s}}{\bar{s}}(x)$.

Within the measured uncertainties, and within the measured $x$-region, the valence quarks ($u$ and $d$) are
polarized and the sea quarks ($\bar{u}$, $\bar{d}$, and $s$) are unpolarized.  The integral value of the measured polarized
strange quark distribution is
$$``\Delta s\mbox{''} = \int_{x=0.023}^{0.30} \Delta s(x) dx = +0.03\pm0.03{\rm(stat)}\pm0.01{\rm(syst)}.$$
[Note this would only be the true $\Delta s$ if the integral was over the range $0<x<1$.]  

Given the fact that the inclusive analysis described in the previous section produced a negative value of $\Delta s$, it is
natural to ask ``where did the negative $\Delta s$ go?''  If the analyses shown of the 
inclusive and semi-inclusive data are both correct, then all the negative contribution to the value of $\Delta s$ must come
from the unmeasured $x$-region, that is from $x<0.023$.  That would imply an average value of $\Delta s(x)$ of approximately
$-5$ in the range $x<0.023$, which is not impossible, as $s(x)$ is of order 20-300 in the 
range $0.01<x<0.001$\cite{Pumplin:2002vw}.
Some very interesting physics indeed would be revealed, if the ``turn on'' of the strange quark polarization in the 
low-$x$ region was this dramatic.

Of course, there are other explanations.  The invocation of SU(3)$_f$ symmetry in the analysis of the inclusive data is
known to be problematic.  And the extrapolations to $x=0$ in those analyses do not have firm theoretical support.
It is clear that a {\em direct} measurement of $\Delta s$ would serve to clarify these issues.

\section{Parity-violating $\vec{e}N$ Elastic Scattering}

One of the highlights of the current low- and medium-energy electron scattering program is the measurement of the
strange vector form factors of the nucleon via parity-violating $\vec{e}N$ scattering.  These measurements are
sensitive as well to the non-strange part of the axial form factor, but rather insensitive to the strange axial form
factor due to the relative sizes of kinematic factors multiplying the various form factors that contribute to the asymmetry.
To be specific, the parity-violating asymmetry observed in these
experiments, when the target is a proton, can be expressed as\cite{Musolf:1994tb}
\begin{eqnarray*}
A_p & = &
\left[\frac{-G_F Q^2}{4\pi\alpha\sqrt{2}}\right]  \\
  & \times &
\frac{\epsilon G_E^\gamma G_E^Z + \tau G_M^\gamma G_M^Z - (1-4\sin^2\theta_W)\epsilon' G_M^\gamma G_A^e}
     {\epsilon\left(G_E^\gamma\right)^2 + \tau\left(G_M^\gamma\right)^2}
\end{eqnarray*}
where $G_{E(M)}^\gamma$ are the traditional electric (magnetic) form
factors of the proton and $G_{E(M)}^Z$ are their weak
analogs, $\tau=Q^2/4M_p^2$, $M_p$ is the mass of the proton,
$\epsilon=[1+2(1+\tau)\tan^2(\theta/2)]^{-1}$, $\theta$ is the
electron scattering angle, and
$\epsilon'=\sqrt{\tau(1+\tau)(1-\epsilon^2)}$.  Lastly, $G_A^e$ is the
effective axial form factor seen in electron scattering:
$$ G_A^e = -G_A^{CC}(1+R_A^{T=1}) + G_A^s + R_A^{T=0}.$$ 
Here, $G_A^{CC}$ is the non-strange (CC) axial form factor, $G_A^s$ is
the strange axial form factor, and the terms $R_A^{T=0,1}$ 
represent electroweak radiative corrections\cite{Musolf:1994tb,Musolf:1990ts,Musolf:1992xm,Zhu:2000gn}.  
The presence of these radiative corrections 
clouds the interpretation of the axial term extracted from these experiments.  To solve this problem, the
SAMPLE\cite{SAMPLE} experiment measured also the same asymmetry on a deuterium target, in which case the
relative kinematic factors of the non-strange ($T=1$) and strange ($T=0$) parts of the axial form factor 
are changed, allowing a separation of the two.
However, one may show that this does not help in identifying the value of $G_A^s$, because the relative size of
the kinematic factors for $G_M^s$ and $G_A^s$ remain the same for either target:  
$$\frac{\partial G_M^s}{\partial G_A^s} = -(1-4\sin^2\theta_W)\frac{\epsilon'}{\tau}\approx-\frac{1}{2}~{\rm for~SAMPLE}.$$
Therefore, parity-violating $\vec{e}N$ scattering experiments can only establish a relationship between the strange
magnetic and axial form factors, they cannot measure them separately.

\section{Combining $\nu N$ and parity-violating $\vec{e}N$ elastic data}

Recently it has been demonstrated\cite{Pate} that the best (perhaps only) way to obtain all three strange quark form factors
(electric, magnetic and axial) is through a combination of low energy $\nu N$ and partity violating $\vec{e}N$
elastic scattering.  This prescription is briefly summarized here.  Using the known values for the electric,
magnetic and non-strange (CC) axial form factors of the proton and neutron, one may take the difference of the $\nu p$ and
$\bar{\nu} p$ elastic cross sections and show it to be a function only of the strange magnetic and axial form factors, 
$G_M^s$ and $G_A^s$.  At the same time, the sum of the $\nu p$ and $\bar{\nu} p$ elastic cross sections can be shown to be
a function only of the strange electric and magnetic form factors, $G_E^s$ and $G_M^s$. Measurements of forward-angle 
parity-violating $\vec{e}p$ elastic scattering are largely functions only of $G_E^s$ and $G_M^s$ as well.  Therefore,
combining these three kinds of data can determine all three strange form factors.  At the present time, there is only
sufficient data at $Q^2=0.5$~GeV$^2$ to make such a determination.  In Ref~\cite{Pate} it is shown that by 
combining the E734\cite{BNL734} results with the HAPPEX\cite{HAPPEX} data from JLab, there are two
possible solutions at $Q^2=0.5$~GeV$^2$, as summarized in Table~\ref{sff_table}.  
\begin{table}[t]
\caption{\label{sff_table} Two solutions for the strange form
  factors at $Q^2=0.5$~GeV$^2$ produced from the E734 and HAPPEX data. (Table from Ref.~\protect\cite{Pate}.)}
\begin{center}
\begin{tabular}{c|c|c}
      & Solution 1 & Solution 2 \\
\hline
$G_E^s$ & $~~0.02\pm 0.09$ & $~~0.37\pm 0.04$ \\ 
$G_M^s$ & $~~0.00\pm 0.21$ & $ -0.87\pm 0.11$ \\
$G_A^s$ & $ -0.09\pm 0.05$ & $~~0.28\pm 0.10$
\end{tabular}
\end{center}
\end{table}
In the long run, additional
experimentation (already on the schedule at JLab) will select one of these solution sets, but there are
already several good reasons to favor Solution~1 over Solution~2.  The value of the strange electric form factor $G_E^s$
must approach zero as $Q^2\rightarrow0$, and Solution~1 is consistent with that requirement.
The value of $G_A^s$ in Solution~1 is consistent with the
estimated value of $G_A^s(Q^2=0) = \Delta s \approx -0.1$ from
deep-inelastic data, whereas that found in
Solution~2 is much larger and of a different sign.  The
value of $G_M^s$ in Solution~1 is consistent with that measured by
SAMPLE\cite{SAMPLE} at $Q^2=0.1$~GeV$^2$, whereas the value in Solution~2 is much larger in magnitude.  Finally, the
value of $G_M^s$ in Solution~1 is consistent with the value of $G_M^s(Q^2=0)= -0.051\pm 0.021 \mu_N$ predicted recently from
lattice QCD\cite{Leinweber:2004tc}.  It seems nearly certain that Solution~1 is the true physical solution.
Future experimentation will in any event select the correct solution set.

Additional data from the $G^0$ Experiment\cite{G0}, recently collected and in the process of analysis, 
will allow the extraction of the three strange form factors when combined with the E734 data.  However,
it is unlikely that knowledge of the strange axial form factor over the range $0.5<Q^2<1.0$~GeV$^2$ will prove
sufficient for the extrapolation to $Q^2=0$ needed for a determination of $\Delta s$.
New neutrino data are needed at lower $Q^2$ to permit a good determination of $\Delta s$.

\section{A future experiment to measure the strange axial charge}

Even if the program I have described determines the strange axial form
factor down to $Q^2$ = 0.45 GeV$^2$ successfully, it almost certainly 
will not determine the $Q^2$-dependence sufficiently for an extrapolation 
down to $Q^2$ = 0. Also, questions remain about the normalization of the 
E734 data. Most of their target protons were inside of carbon nuclei, and 
there was not much known about nuclear transparency in the mid-1980's.  
The E734 collaboration did make a correction for transparency effects, 
but this issue needs to be revisited if we continue to use the E734 data.

A new experiment\cite{FINeSSE} has been proposed to measure elastic and quasi-elastic 
neutrino-nucleon scattering to sufficiently low $Q^2$ to determine the
strange axial charge, $\Delta s$. The FINeSSE Collaboration proposes to measure the NC to CC neutrino scattering ratio
$$R_{NC/CC}=\frac{\sigma(\nu p \rightarrow \nu p)}{\sigma(\nu n \rightarrow \mu^- p)}$$
and from it extract the strange axial form factor down to $Q^2=0.2$~GeV$^2$.  The numerator in this ratio
is sensitive to the full axial form factor, $-G_A^{CC} + G_A^s$, while the denominator is sensitive to only $G_A^{CC}$.
The processes in the numerator and denominator have unique charged particle final state signatures.
This ratio is largely insensitive to
uncertainties in neutrino flux, detector efficiency and nuclear target effects. A 6\% measurement of $R_{NC/CC}$
down to $Q^2$ = 0.2 GeV$^2$
provides a $\pm0.04$ measurement of $\Delta s$.
\begin{acknowledgement}
This work was supported by the US Department of Energy.
\end{acknowledgement}

\end{document}